\documentclass[floatfix,twocolumn,showpacs]{revtex4}
\usepackage{graphicx}
\usepackage{tikz}
\usepackage{multirow}
\usepackage{dcolumn}
\usepackage{bm}
\usepackage{epsfig}
\usepackage{floatrow}
\newcommand{\be}{\begin{equation}}
\newcommand{\ee}{\end{equation}}
\newcommand{\ba}{\begin{eqnarray}}
\newcommand{\ea}{\end{eqnarray}}
\newcommand{\etal}{{\em et al.}}

\begin{document}
% terminat comenzile din prc9.tex

% ****** Start of file apssamp.tex ******
%
%   This file is part of the APS files in the REVTeX 4 distribution.
%   Version 4.0 of REVTeX, August 2001
%
%   Copyright (c) 2001 The American Physical Society.
%
%   See the REVTeX 4 README file for restrictions and more information.
%
% TeX'ing this file requires that you have AMS-LaTeX 2.0 installed
% as well as the rest of the prerequisites for REVTeX 4.0
%
% See the REVTeX 4 README file
% It also requires running BibTeX. The commands are as follows:
%
%  1)  latex apssamp.tex
%  2)  bibtex apssamp
%  3)  latex apssamp.tex
%  4)  latex apssamp.tex
%
%\documentclass[twocolumn,showpacs,preprintnumbers,amsmath,amssymb]{revtex4}
%\documentclass[preprint,showpacs,preprintnumbers,amsmath,amssymb]{revtex4}

% Some other (several out of many) possibilities
%\documentclass[preprint,aps]{revtex4}
%\documentclass[preprint,aps,draft]{revtex4}
%\documentclass[prb]{revtex4}% Physical Review B

%\usepackage{graphicx}% Include figure files
%\usepackage{dcolumn}% Align table columns on decimal point
%\usepackage{bm}% bold math

%\nofiles

%\begin{document}

%\preprint{APS/123-QED}

\title{Peripheral elastic and inelastic scattering  of $^{17,18}$O on light targets at 12 MeV/nucleon}

\author{T. Al-Abdullah$^{1,3}$, F. Carstoiu$^2$\footnote{carstoiu@theory.nipne.ro}, C.A. Gagliardi$^1$, G. Tabacaru$^1$, L. Trache$^{1,2}$ and  R.E. Tribble$^1$}
% \altaffiliation[current address]{Physics Department, The Hashemite University. Zarqa, Jordan}%Lines break
% automatically or can be forced with \\

%\author{X. Chen, H.L. Clark, C.A. Gagliardi, Y.-W. Lui,\\ A. Mukhamedzhanov, G. Tabacaru, Y. Tokimoto, L. %Trache, R.E. Tribble, and Y. Zhai}
%\affiliation{Cyclotron Institute, Texas A$\&$M University, College Station, Texas 77843-3366}

\affiliation{
$^1$Cyclotron Institute, Texas A$\&$M University, College Station, Texas 77843, USA\\
$^2$National Institute for Physics and Nuclear Engineering Horia Hulubei, Bucharest, Romania\\
$^3$Physics Department, The Hashemite University. Zarqa, Jordan}
\date{\today}% It is always \today, today,

\begin{abstract}

A study of interaction of neutron rich oxygen isotopes $^{17,18}$O with light targets
has been undertaken in order to determine the optical potentials needed for the transfer
reaction $^{13}$C($^{17}$O,$^{18}$O)$^{12}$C. Optical
potentials in both incoming and outgoing channels have been determined in a single
experiment. This transfer reaction was used to infer the direct capture rate to the
$^{17}$F(p,$\gamma$)$^{18}$Ne which is essential to estimate the production of
$^{18}$F at stellar energies in ONe novae. The success of the asymptotic normalization
coefficient (ANC) as indirect
method for astrophysics is guaranteed if the reaction mechanism is peripheral
and the DWBA cross section calculations are warranted and stable against OMP used.
We demonstrate the stability of the ANC method and OMP results
using good quality elastic and inelastic scattering data with stable beams  before
extending the procedures to rare ion beams.
The peripherality of our reaction is inferred from a semiclassical decomposition
of the total scattering amplitude into barrier and internal barrier components.
 Comparison between elastic
scattering of $^{17}$O, $^{18}$O and $^{16}$O projectiles is made.
\end{abstract}

\maketitle

\section{Introduction}
The $^{17}$F(p,$\gamma$)$^{18}$Ne reaction is important for understanding nucleosynthesis
in novae and plays a role in determining if
radioactive nuclei with characteristic gamma-ray signature are produced in sufficient
yield to be observed by gamma-ray satellites.
 The reaction rate is expected to be dominated by direct-capture cross section at nova
temperatures and influences the abundances of $^{15}$O,
$^{17}$F, $^{18}$F and $^{18}$Ne \cite{garcia}. The rate also determines the $^{17}$O/$^{18}$O ratio that is
produced and explains the transition sequence
from the HCNO cycle to the {\it  rp}-process \cite{wallace}.

%Nucleosynthesis of elements in ONe white dwarf (WD) novae produces several sources
%of $\gamma$-ray lines. Among them is the positron-electron annihilation in the
%nova envelope, which leads to emission of a line at 511 keV and a continuum
%below it.  The decay of $^{18}$F is
%important since its $\gamma$-ray photons are emitted when the envelope starts to be
%transparent. According to the ONe models, when the temperature in the burning shell
%reaches T$_{9}\sim 0.2-0.4$, the main nuclear activity to produce $^{18}$F is driven
%by a $\beta$-decay following the proton capture reaction $^{17}$F($p$,$\gamma$)$^{18}$Ne \cite{Coc2000}.

%The nuclear structure of $^{18}$Ne is related to the configurations and the
%excitation spectrum in the mirror nucleus $^{18}$O.
%Shell model calculations assume a 2$s$ or 1$d$ nucleon
%coupled to the single particle 5/2$^ +$, 1/2$^ +$, and
%3/2$^ +$ levels of $^{17}$O and $^{17}$F. Comparison of the nuclear structure of the
%mirror nuclei for the low-lying states (${2_1^ +  }$,${4_1^ +  }$,${0_2^ +  }$,${2_2^ +  }$,${2_3^ +  }$,${3_1^ +  }$)
%shows that their excitation energies are very similar as reported in \cite{Sherr1998}.

The importance of the direct capture to the bound states in $^{18}$Ne has  been
recently estimated by our team \cite{tariq2013}. Because of the difficulties of
obtaining information from experiments
with radioactive beams,  the asymptotic normalization coefficients (ANCs) as an
alternative technique to determine this direct capture reaction rate has been used. The
spectroscopic factors for the major components of the lowest
lying states in
mirror nuclei are the same, so the ANC method can be applied to the mirror nucleus $^{18}$O and used
to extract the ANCs for the g.s., $E_x \left( {2_1^ +  } \right) = 1.982$ MeV
and $E_x \left( {2_2^ +  } \right) = 3.920$ MeV
states and convert them to their corresponding states in $^{18}$Ne. The primary goal of the
 experiment was the measurement of the peripheral
neutron transfer reaction $^{13}$C($^{17}$O,$^{18}$O)$^{12}$C. Optical
potentials in the incoming and outgoing
channels have been obtained by measuring elastic scattering angular
distributions $^{17}$O+ $^{13}$C and $^{18}$O+
$^{12}$C at 12 MeV/nucleon incident energy. The quality of the obtained
potentials has been also checked from inelastic scattering to selected states in
$^{17}$O$^*$ and $^{18}$O$^*$. Since the ANC method assumes the peripherality of
the reaction mechanism, we discuss here rather extensively this issue by
decomposing semiclassically the total scattering amplitude into barrier and
internal barrier subcomponents. We show that the internal barrier subcomponent,
which corresponds to the flux penetrating the barrier, gives negligible small
contribution to the total cross section, and thus the reaction is peripheral.
The elastic scattering $^{17}$O+$^{13}$C includes a fragile target. A difficulty
in obtaining the optical model parameters in this type of
reactions may arise due to the competition between the increased refractive
power of the real potential and increased absorption at the nuclear
surface. The well known existence of many ambiguities in the optical model parameters
extracted from elastic scattering can raise questions about the reliability
and accuracy of these determinations.

Previously, $^{18}$O+$^{12}$C elastic scattering at barrier energies was measured
by Robertson \etal \cite{robertson},
 by Szilner \etal \cite{szilner} and Rudchik \etal \cite{rudchik} at some
5-7 MeV/nucleon. Fresnel scattering of $^{18}$O on $^{28}$Si was measured by
Mermaz \etal  \cite{mermaz} at 56 MeV. For the $^{17}$O+ $^{13}$C reaction
the data are rather scarce, we identified a single fusion study and poor elastic angular
distributions at barrier energies \cite{heusch}.
The main conclusion of these studies was that the interaction of $^{17,18}$O nuclei
with light targets is slightly more
absorptive compared with that of the closed shell nucleus $^{16}$O and that no significant
effects due to the neutron excess were
identified.

In Sec. II we give a short description of the experiment. Elastic scattering data and the derivation of the OM potentials
are discussed in Sec. III. The semiclassical (WKB) method is used in Sec IV to decompose the total scattering amplitude into
barrier and internal barrier components. Inelastic angular distributions to selected states in $^{18}$O$^*$ and $^{17}$O$^*$ are
discussed in Sec. V. Our conclusions are summarized in Sec. VI.

\section{The experiment}
The primary goal of the experiment was the measurement of the transfer reaction
$^{13}$C($^{17}$O,$^{18}$O)$^{12}$C at 12 MeV/nucleon. In addition, elastic scattering in
both incoming
and outgoing channels as well as inelastic scattering to selected states in
$^{17}$O and $^{18}$O were measured.

 The experiment was carried out with two separate
 $^{17}$O and $^{18}$O beams from K500
superconducting cyclotron at Texas A$\&$M University. Each beam was transported through the
beam analysis system to the scattering chamber of the multipole-dipole-multipole (MDM) magnetic
spectrometer \cite{MDM}, where it interacted with 100 $\mu$g/cm$^2$
self-supporting targets.

 First, the $^{17}$O beam  impinged on $^{13}$C target. The elastic
scattering angular distribution was measured for the spectrometer angles 4$^\circ$-25$^\circ$ in the laboratory system.
Fine tuned RAYTRACE \cite{Raytrace} calculations were used to reconstruct the
position of particles in the focal plane and the scattering angle at the target.
A $4^\circ \times 1^\circ$ wide-opening mask and an angle mask consisting of five narrow ($\Delta\theta$ = 0.1$^\circ$)
slits were used for each spectrometer angle to double-check the absolute values of the cross section and the
quality of the angle calibration.  The instrumental setup,
including the focal plane detector, and processes for energy and angle calibrations, are identical to that
described in Ref. \cite{Akram1997}. Second, the $^{12}$C target was bombarded by $^{18}$O beam with 216 MeV
total laboratory energy. The elastic scattering cross section was measured at 4$^\circ$-22$^\circ$ spectrometer angles.

\begin{figure}[!ht]
\includegraphics[trim= 0.0cm 0.0cm 0.0cm 0.0cm , clip=true,width=\textwidth]{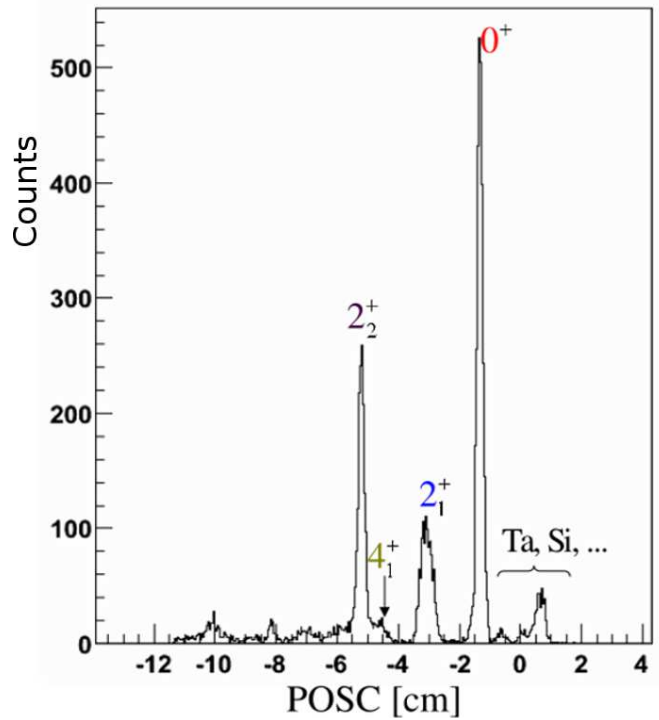}
\caption{\label{figtar1}(Color online) Low-lying spectrum of $^{18}$O versus
the particle position in the focal plane, measured at the spectrometer angle of 4$^\circ$. The peaks
at the right of the  elastic peak are due to Si and Ta contaminants in the target holder.}
\end{figure}
The angular resolution, $\Delta \theta_{res}$, of the detector in both cases was on
average 0.31$^\circ$ in c.m. frame and the position resolution was better than 1 mm. The low lying spectrum of $^{18}$O
as a function of the position in the focal plane is shown in Fig.\ref{figtar1}. The spectrum is taken
at the spectrometer angle of 4$^\circ$. The peaks corresponding to elastic
scattering and to inelastic transitions to
the $2_1^+$ and $2_2^+$ excited states were observed with sufficient statistics
over the whole angular range to obtain good angular distributions. Small amounts of
heavy impurities in the target most likely Ta and Si dominate
the spectrum at small angles ( bellow $\theta_{lab}=3^{\circ}$). The absolute values of the cross section
were determined by a careful integration of beam charge in a Faraday cup and the
measurement of target thickness from the energy loss using alpha particles from sources
and  the beam. The overall normalization of data was also extensively checked by
comparing the data
at the most forward angles with optical model calculation. At these angles the cross section is less sensitive to the
nuclear potential. The main uncertainties in the data are due to 7.5\% in the target thickness and 3\%
statistical errors. The average normalization error was less than 3$\%$.

\begin{figure}[!ht]
{\includegraphics[trim= 1.0cm 0.0cm 1.1cm 1.0cm , clip=true, scale=0.45]{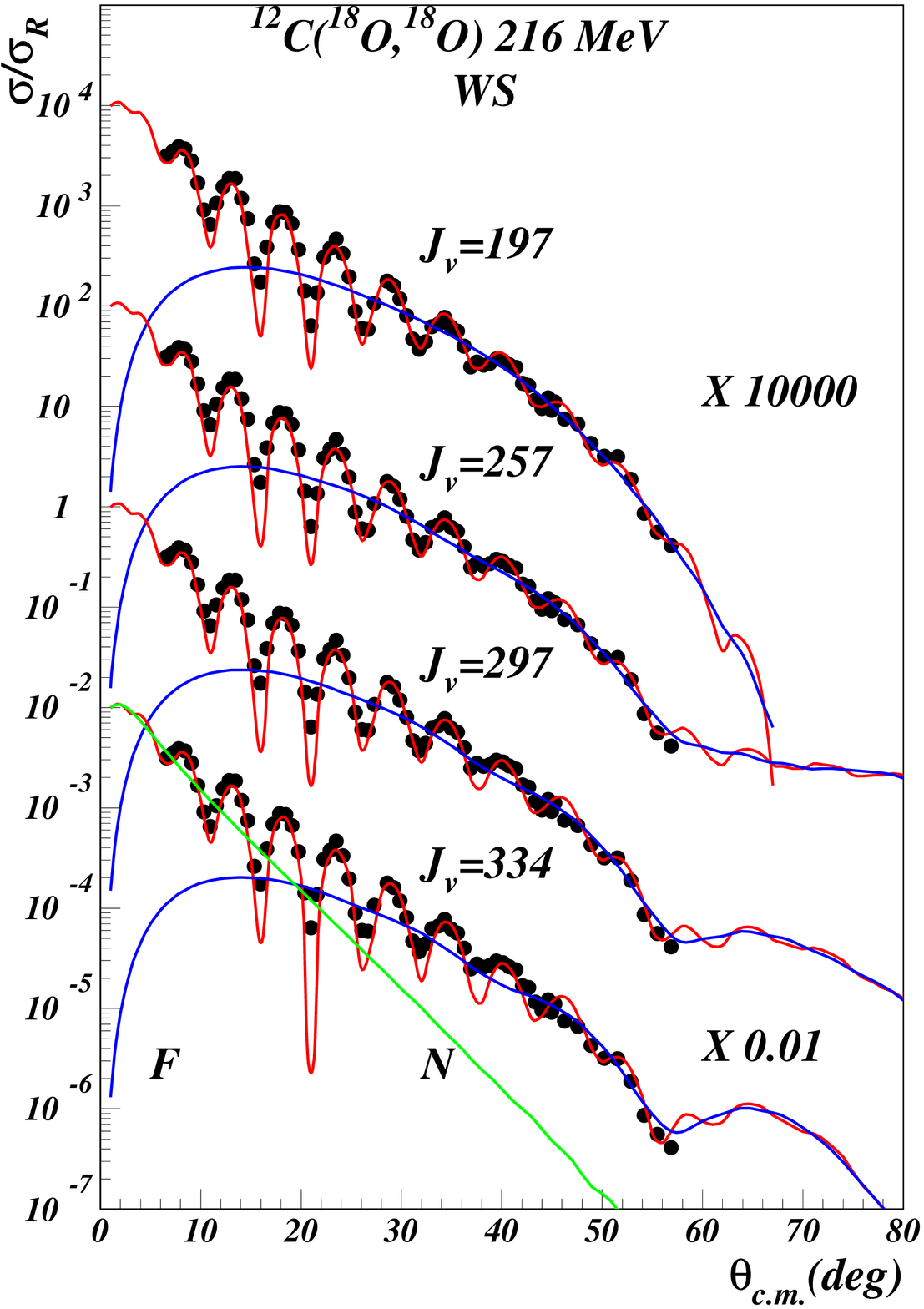}}{\caption{
(Color online) Cross section and far side/near side (F/N) decomposition of the scattering amplitude for WS potentials in Table \ref{tableo18c12216mevws}. Each calculation is identified by its real volume integral $J_V$ and shifted by factors X to increase the visibility.}\label{figtar2}}
\end{figure}

\begin{figure}[!ht]
{\includegraphics[trim= 1.0cm 0.0cm 1.1cm 1.0cm , clip=true, scale=0.45]{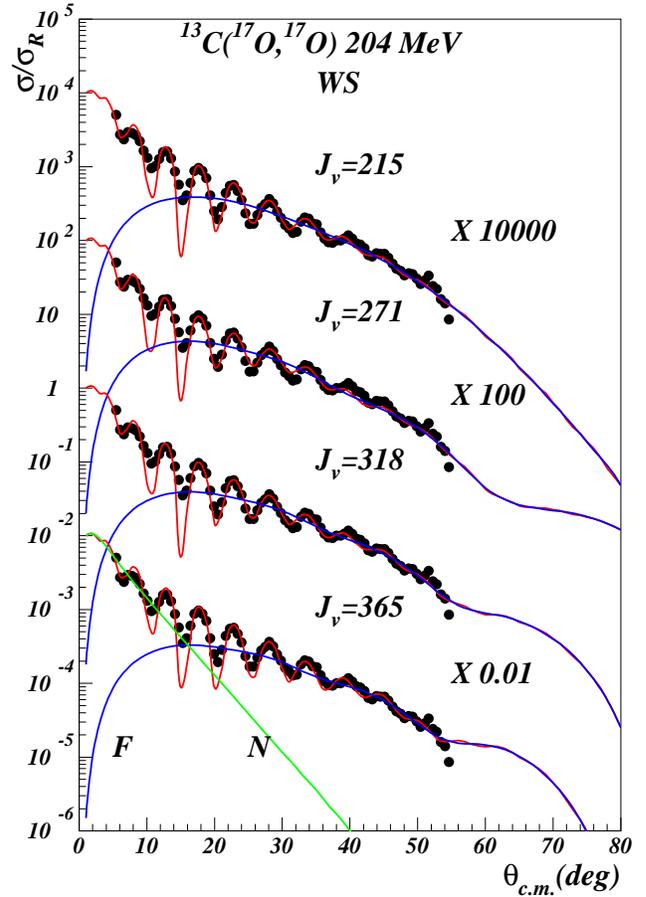}}{\caption{(Color online) Cross sections and
F/N decomposition for the WS potentials Table \ref{tableo18c12216mevws}.
The far side component shows Airy oscillation which moves to forward angles with
increased value of the real volume integral.}\label{figtar3}}
\end{figure}

% large fonts executed
\begin{figure}[!ht]
{\includegraphics[trim= 1.0cm 1.0cm 1.1cm 2.0cm , clip=true, scale=0.5]{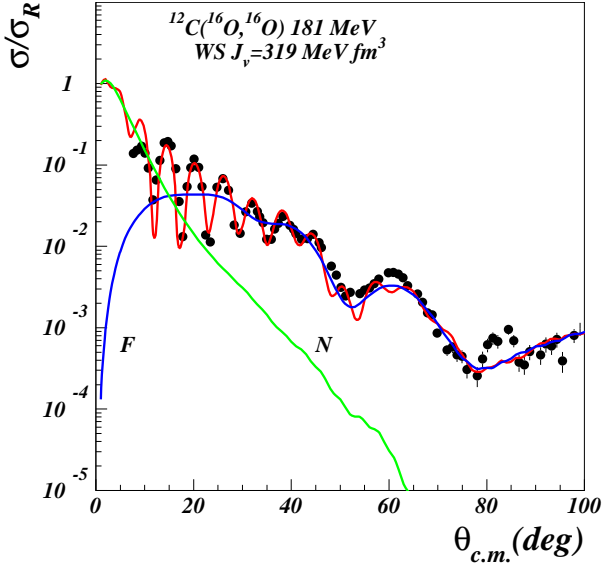}}{\caption{ (Color online) Elastic
scattering $^{16}$O+$^{12}$C at 11.3 MeV/nucleon. The real part of the WS optical potential is much
stronger and the far side component shows several deep Airy oscillations. Experimental data are taken from \cite{glukhov}.}\label{figtar4} }
\end{figure}

\begin{figure}[!ht]
{\includegraphics[trim= 1.0cm 0.0cm 1.1cm 1.0cm , clip=true, scale=0.5]{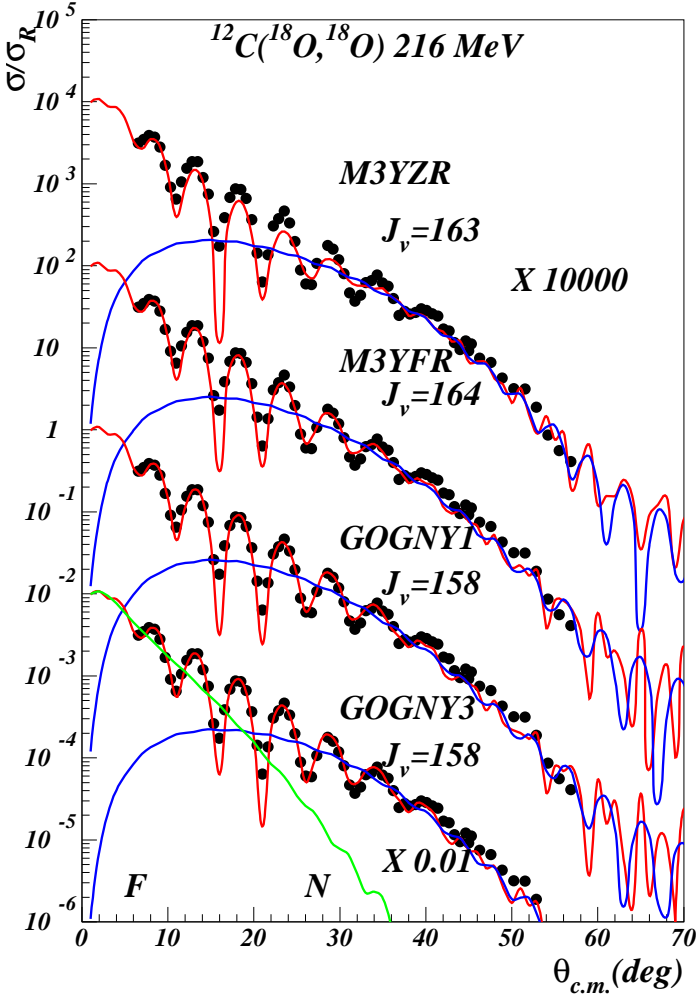}}{\caption{(Color online) Cross
section calculated with folding form factors using the M3Y and GOGNY models. The real volume integral
is indicated on each curve. The far side/near side components of the cross section are denoted by F/N. Experimental data
and calculation have been shifted by factor X to increase visibility.}\label{figtar5}}
\end{figure}

\begin{figure}[!ht]
{\includegraphics[trim= 1.0cm 0.0cm 1.1cm 1.0cm , clip=true, scale=0.5]{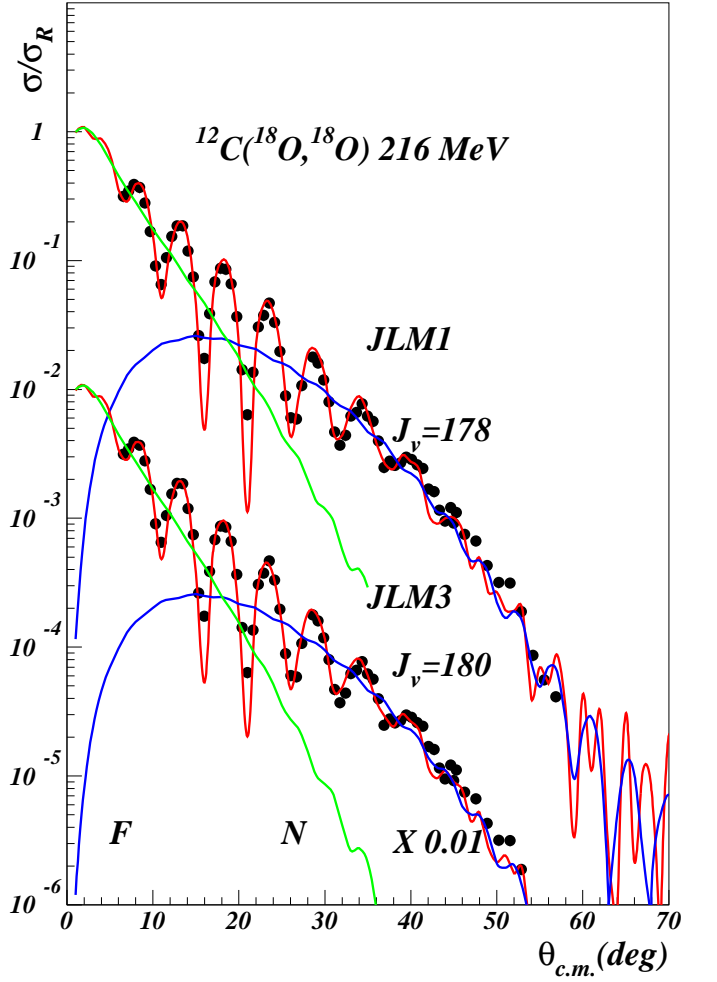}}{\caption{(Color online) The same
as in Fig.\ref{figtar5} but for the JLM model.}\label{figtar6}}
\end{figure}
\begin{table*}
  \centering
  \caption{\label{tableo18c12216mevws}
Discrete solutions obtained with WS form factors for $^{18}$O+$^{12}$C at 216 MeV and $^{17}$O+$^{13}$C at 204 MeV.
The line labeled PP9 is a WS phase equivalent of the JLM1 solution.}
\begin{ruledtabular}\begin{tabular}{ccccccccccccc}
pot&V	&	W	&	$r_V$&	$r_W$	&$a_V$&	$a_W$&	 $\chi^2$&$\sigma_R$&	$J_V$&	$R_V$&	$J_W$&	 $R_W$ \\
   & MeV& MeV           &       fm   &   fm     &  fm & fm   &          &  mb     &MeV fm$^3$& fm    &MeV fm$^3$&  fm\\
\multicolumn{13}{ c }{$^{18}$O+$^{12}$C at 216 MeV} \\

PP5 &  89.18	&25.24&	0.88&	1.16&	0.88&	0.68&	5.12	 &1712&	197&	4.69	&103&	5.09 \\
PP6 & 195.40&	25.59&	0.68&	1.16&	0.96&	0.67&	6.39	 &1702&  257&	4.40	&104&	5.07 \\
PP7 & 295.82&	26.00&	0.60&	1.16&   0.95&	0.67&	7.54    &1696&	297&    4.20    &106&   5.06 \\
PP8 & 374.41&	26.19&	0.58&	1.16&	0.90&	0.68&	9.78	 &1695&	334&	4.01	&107&	5.06 \\
PP9 &  75.68&   26.16&  0.89&   1.15&   0.93&   0.66&   5.31    &1677&  178&    4.85    &104&   5.02\\
  &&&&&&&&&&&&\\
  \multicolumn{13}{ c }{$^{17}$O+$^{13}$C at 204 MeV} \\

T1  &   94.69&	26.91&	.91&	1.13&   .84&	.67 &   4.47&	 1659&	215&    4.67&	99&	4.96\\
T2  &  188.40&	24.95&	.72&	1.12&	.94&	.69 &	4.62&	 1667&	271&	4.44&	92&	4.99\\
T3  &  248.75&	26.36&	.69&	1.13&	.90&	.66 &	4.53&	 1659&	318&	4.27&	99&	4.97\\
T4  &  275.49&  25.63&	.73&	1.15&	.81&	.65 &	5.90&	 1660&	365&	4.11&  100&	5.00\\
\end{tabular}\end{ruledtabular}
\end{table*}

\begin{table*}
  \centering
  \caption{\label{tableo18c12216mevfolding}
Unique solutions obtained with folding form factors for $^{18}$O+$^{12}$C at 216 MeV and $^{17}$O+$^{13}$C at 204 MeV.}
\begin{ruledtabular}\begin{tabular}{ccccccccccc}
 pot&$N_V$&	$N_W$	&   $t_V$	&	$t_W$&	 $\chi^2$	&  $\sigma_R$	&    $J_V$&		 $R_V$	&$J_W$	&  $R_W$\\
    &     &             &               &            &                  & mb            &MeV fm$^3$&   fm               &MeV fm$^3$& fm \\
\multicolumn{11}{ c }{$^{18}$O+$^{12}$C at 216 MeV} \\
M3YZR   &    0.37  &  0.20  &  0.88  &  0.80  &   10.72   &  1812 &   163   &  4.60  &   86  &   5.06\\
M3YFR   &    0.33  &  0.21  &  0.88 &   0.86  &    8.15   &  1737 &   164   &  4.68  &  103  &   4.83\\
GOGNY1   &   0.28  &  0.18  &  0.89  &  0.87   &   7.27  &   1707 &   158   &  4.70  &  103  &   4.83\\
GOGNY3  &    0.37  &  0.21  &  0.91 &   0.84 &     7.39   &  1767  &  158   &  4.69  &   89  &   5.08\\
JLM1    &    0.33  &  0.93  &  0.87  &  0.86  &    6.87   &  1675 &   178  &   4.55  &  109   &  4.80\\
JLM3     &   0.36  &  1.02  &  0.86  &  0.85  &    6.75   &  1708  &  180  &   4.56 &   102 &    4.85\\
  &&&&&&&&&&\\
  \multicolumn{11}{ c }{$^{17}$O+$^{13}$C at 204 MeV} \\
 M3YZR  &    0.46 &   0.22 &   0.91 &   0.85 &     5.24  &   1742 &   203  &   4.48  &   95 &    4.80\\
M3YFR &     0.38  &  0.18  &  0.93 &   0.86  &    5.16 &    1738  &  196  &   4.52  &   94  &   4.87\\
GOGNY1 &    0.32  &  0.15  &  0.94 &   0.85  &    5.74 &    1748  &  188   &  4.53  &   88  &   4.99\\
GOGNY3 &    0.41 &   0.20  &  0.95 &   0.87  &    6.03  &   1729 &   186   &  4.53   &  88  &   4.97\\
JLM1   &    0.35  &  0.72  &  0.89  &  0.84  &    6.06 &    1691  &  196   &  4.47  &   84  &   4.96\\
JLM3   &    0.37  &  0.80  &  0.88  &  0.83  &    5.63  &   1719 &   192  &   4.49  &   81 &    5.00\\
\end{tabular}\end{ruledtabular}
\end{table*}

\section{Elastic scattering}
\subsection{Woods-Saxon formfactors}
The measured elastic scattering data at $E_{lab}$=216 and 204 MeV are shown in
Figs. \ref{figtar2} and \ref{figtar3}.
  The data are first analyzed using optical potentials
with conventional Woods-Saxon (WS) form factors for the nuclear term,
supplemented with a Coulomb potential generated by a uniform charge
distribution with a reduced radius fixed to $r_{c}$=1 fm. No preference has
been found for volume or surface localized absorption and throughout the
paper only volume absorption is considered. In the absence of any spin
dependent observables, spin-orbit or tensor interactions have been ignored.
Ground state reorientation couplings  have been neglected also. The potential
is defined by six parameters specifying the depth and geometry of the real
and imaginary terms, with the standard notations, the same as used in Ref.
\cite{trache00}. The number of data points N is
quite large, and consequently the usual goodness of fit criteria ($\chi ^{2}$)
normalized to  N has been used.

Using the strength of the real component of the optical potential as a control
parameter, a grid search procedure revealed a number of discrete solutions.
Their parameters are presented
in Table \ref{tableo18c12216mevws}. All of the potentials give relatively
small $\chi^2$, but only those with the
smallest values for entrance and exit channels, potential T1 and PP5 respectively,
were adopted in the DWBA calculations
of the neutron transfer reaction \cite{tariq2013}, while the others were used to determine
the uncertainty in the choice of the OMP
in either channel. The ambiguity in the optical potential has two main
sources: the limited range of the
measured angles and the strong absorption. When the strong absorption
dominates the reaction mechanism,
then the interaction is sensitive only to the surface and several phase
equivalent optical potentials  will appear. The patterns shown
in Figs \ref{figtar2} and \ref{figtar3} show rapid oscillation at
forward angles followed by a smooth fall-off
at intermediate angles. Assuming pure Fraunhofer scattering at
forward angles , we extract a grazing
angular momentum $\ell_g\approx 36$ from the angular
spacing $\Delta\theta=\pi/(\ell_g+1/2)$.
The corresponding grazing distance is quite large, $R_g\approx 7$ fm,
much larger than the distance of touching configuration.
We systematically find  diffuse real potentials ($a_V\approx 0.9$ fm). This effect
may be tentatively attributable to the neutron excess. We find also quite constant  volume integrals
and $rms$ radii for the imaginary component. As a consequence the
total reaction cross section seems to be a well defined observable. Weighted
average values from table \ref{tableo18c12216mevws} and table \ref{tableo18c12216mevfolding} are
$\sigma_R=1713\pm 35$ mb and $\sigma_R=1699\pm 36$ mb
for $^{18}$O+$^{12}$C and $^{17}$O+$^{13}$C reactions respectively. The larger the real volume
integral the smaller reduced radius $r_V$ is
required to match the data and the far-side component becomes
more structured. For the largest real volume
integral an Airy oscillation forward to
a primary rainbow becomes  apparent. Usually, the dominance of the
far-side component beyond the Fraunhofer crossover is
interpreted as a signature of refractive effects due to a strongly
attractive real potential and weak absorption. We will show bellow
that the strong absorption is still the dominant reaction mechanism.

A comparison with the scattering of the tightly bound nucleus $^{16}$O is in
order. Experimental
data \cite{glukhov} and our calculation for $^{16}$O+$^{12}$C at 11.3 MeV/nucleon
are displayed in Fig. \ref{figtar4}.
We did not find any
reasonable WS solution with $J_V<300$ MeV fm$^3$ and so the solution with the
lowest acceptable real volume integral is plotted. Since the potential is strong,
the far-side component of the cross section is
much more structured. While the Fraunhofer (diffractive) part at forward angles
is similar to our reactions, strong refractive
effects appear at $\theta>40^{\circ}$ as deep Airy oscillations.

%folding
\subsection{Folding formfactors}
In the  following we discuss the ability of the folding model
to describe our data. We start by a quite simple model in which the spin-isospin
independent formfactor of the OMP is given by the
double folding integral,
\be
V_{fold}(R)=\int d \vec r_1 d \vec r_2 \rho_1(r_1) \rho_2(r_2)v_{M3Y}(s)
\label{eqff1}
\ee
where $v_{M3Y}$  is the M3Y
parametrization of the G-matrix obtained from the Paris NN interaction \cite
{m3y}, and $\vec s=\vec r_1+\vec R-\vec r_2$ is the NN separation distance. For the
reaction $^{17}$O+$^{13}$C we add the small
isovector component arising from the nonnegligible neutron skin present in both
interacting partners. The Coulomb component of the optical potential is
calculated by replacing the nuclear s.p. densities with proton densities and
using $v_{coul}(s)=e^2/s$ as effective interaction. The small effect arising
from finite proton size is ignored.
 In the simplest
version of this model, dubbed here as M3YZR, the
knockon exchange  component is simulated by a zero range potential
with a slightly energy dependent strength,
\be
J_{00}(E)=-276(1-0.005E/A)
\label{eqff2}
\ee

 We keep the number of fitting parameters at the minimum level and take  the OMP in the form,
\be
U(R)=N_V V(R,t_V)+iN_W V(R,t_W)
\label{eqff3}
\ee
where $N_{V,W}$ are normalization constants and $t_{V,W}$ are range parameters
defined by the scaling transformation,
\be
V(R,t)\rightarrow t^3V_{fold}(tR)
\label{eqff4}
\ee
This transformation conserves the volume integral of the folding potential and modifies the radius as,

\be
<R^2>_V=\frac{1}{t^2}<R^2>_{fold}
\label{eqff5}
\ee

Thus the strength of the formfactor is controlled by the parameters
$N_{V,W}$. Note that the transformation in Eq. (\ref{eqff4}) ensures that only the
$rms$ radius of the bare folding potential is changed. This is in line with the original prescription of
\cite{jeuken} which proposed a smearing procedure in terms of a normalized Gaussian function. We found that
the transformation in Eq. (\ref{eqff4}) is more efficient. Based on Eq. (\ref{eqff5}) one may
estimate in an average
way the importance of
the dynamic polarization potential (DPP) and finite range effects. Throughout this paper we use
single particle densities obtained
from a spherical Hartree-Fock (HF+BCS) calculation based on the density
functional of Beiner and Lombard
\cite{beiner}. The obtained $rms$ charge radii are very close to the experimental values \cite{angeli} and
the model predicts a neutron skin $\Delta r=r_n-r_p$ of 0.1, 0.18 and 0.1 fm for $^{13}$C, $^{18}$O, $^{17}$O
respectively.  The calculated neutron $rms$ radii are 2.84 and 2.76 fm for $^{18}$O, $^{17}$O
in good agreement with the values extracted by
Khoa \etal \cite{khoa} from high energy interaction cross section. Note that for the
fragile $^{13}$C ($S_n=4.9$ MeV)  this model predicts a small occupation
probability for the neutron $2s_{1/2}$ level of $v^2_{2s_{1/2}}=0.0016$ but this
has a small influence on the tail of the s.p. density.
%. Note that the minimum $rms$ radius implied by Eq.
%\ref{eqff6} is 3.57 fm. Far-side/near-side decomposition of the scattering
%amplitude reveals the same features: a minimum in the far-side component
%develops at $\theta=65^{\circ}$ which becomes  deeper with the increased real
%volume integral of the interaction.
A more elaborate calculation leads to a nonlocal knockon exchange kernel \cite{exchange},

\begin{figure}[!ht]
{\includegraphics[trim= 1.0cm 0.0cm 1.1cm 1.0cm , clip=true, scale=0.5]{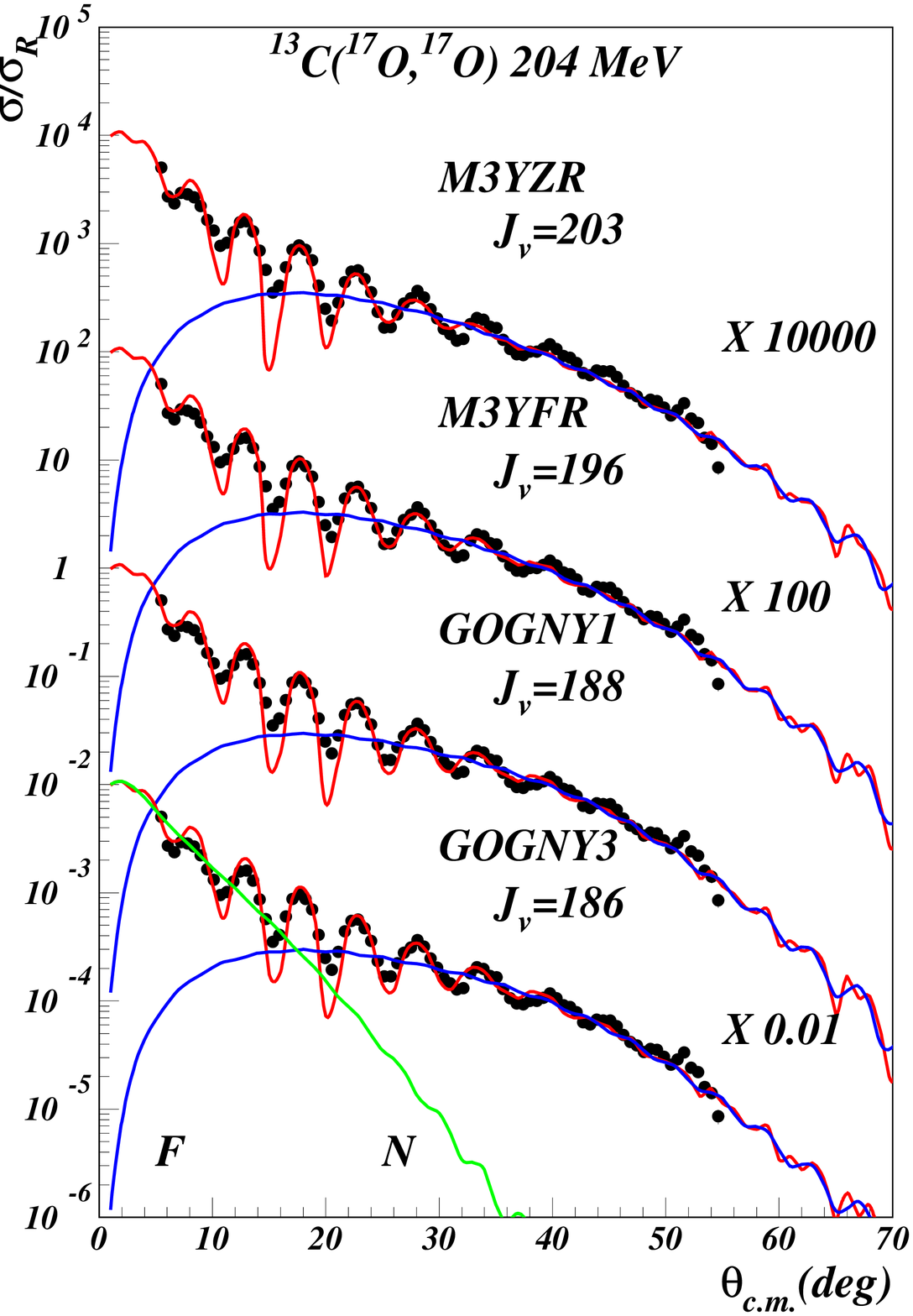}}{\caption{(Color online) Cross
section and F/N decomposition with folding form factors. Parameters are taken from Table \ref{tableo18c12216mevfolding}.}\label{figtar7}}
\end{figure}
\begin{figure}[!ht]
{\includegraphics[trim= 1.0cm 0.0cm 1.1cm 1.0cm , clip=true, scale=0.5]{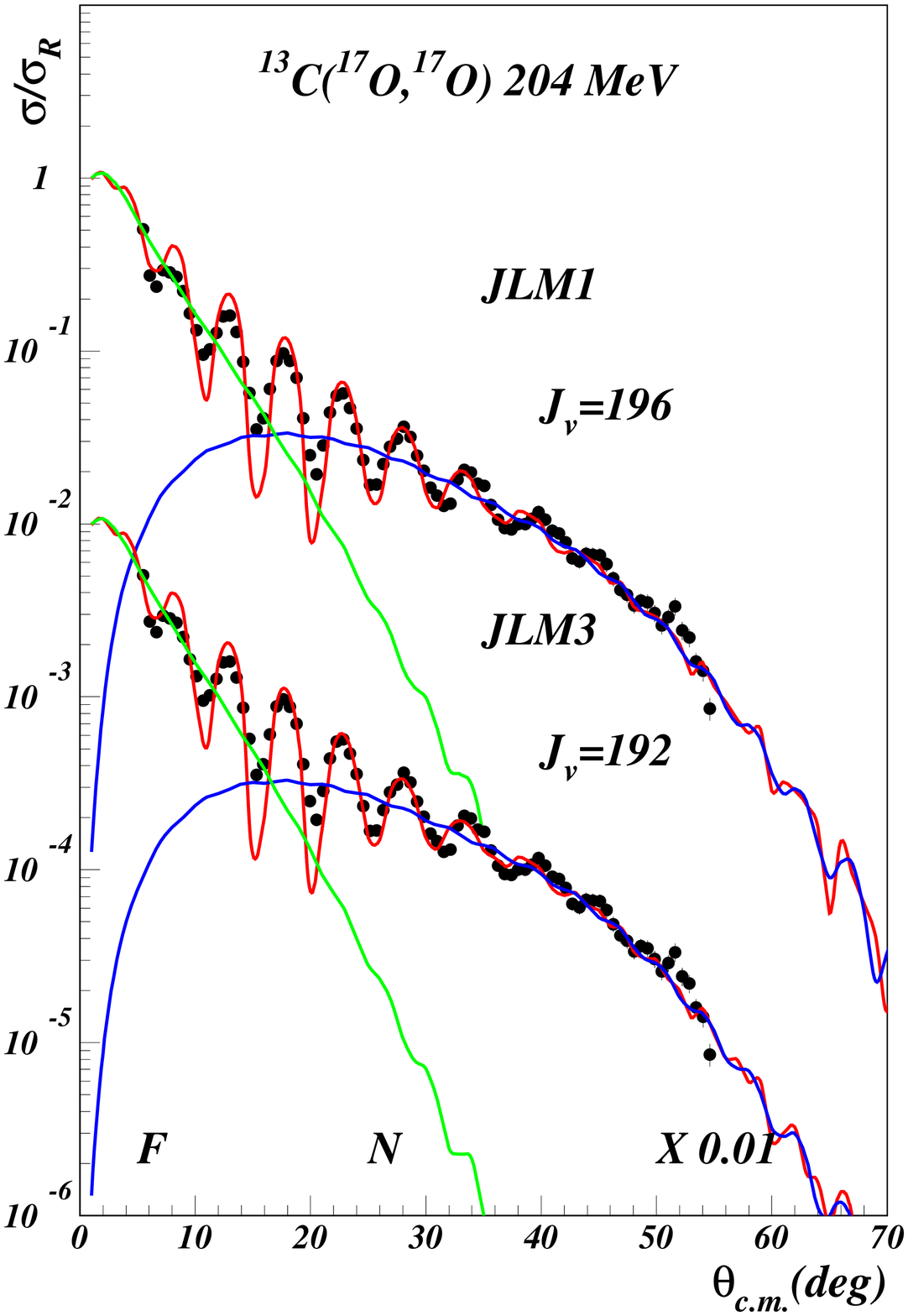}}{\caption{(Color online) Cross section
and F/N decomposition using the JLM form factors.}\label{figtar8}}
\end{figure}
\begin{widetext}
\ba
U_{ex}(\vec R^+,\vec R^-)=\mu^3v_{ex}(\mu R^-)\int d\vec X_1\rho_1(X_1)\hat
j_1(k_{f1}(X_1)\frac{(A_1-1)A_2}{A_1+A_2}R^-)\label{eqf8}\\
\times \rho_2(\vert \vec R^+-\vec X_1\vert)\hat
j_1(k_{f2}(\vert \vec R^+-\vec X_1\vert)\frac{(A_2-1)A_1}{A_1+A_2}R^-)
\nonumber
\ea
\end{widetext}
\noindent
where $A_{1,2}$  are mass numbers, $\mu$ is the reduced mass of the system,
$k_{f1,2}$ are Fermi momenta, $R^{+,-}$ are the usual nonlocal coordinates and
$v_{ex}$ is the exchange component of the interaction including the long range
OPEP tail. Eq. (\ref{eqf8}) already shows that the nonlocality  is small and
behaves as $\sim\mu^{-1}$. In the lowest order of the Perey-Saxon approximation, the local
equivalent of the nonlocal kernel is obtained by solving the nonlinear equation,
\vspace{1cm}
\ba
&&U_L(R)=4\pi\int d\vec r_1 d\vec r_2\rho_1(r_1)\rho_2(r_2) \nonumber \\
&&\times \int s^2 ds v_{ex}(s)\hat j_1(k_{f1}(r_1)\beta_1 s)\hat
j_1(k_{f2}(r_1)\beta_2 s)\nonumber \\
&&\times j_0(\frac{1}{\mu}K(R)s)\delta(\vec r_2-\vec r_1+\vec R)
\label{eqf9}
\ea

\begin{figure}[!ht]
{\includegraphics[trim= 1.0cm 1.0cm 1.1cm 1.0cm , clip=true, scale=0.5]{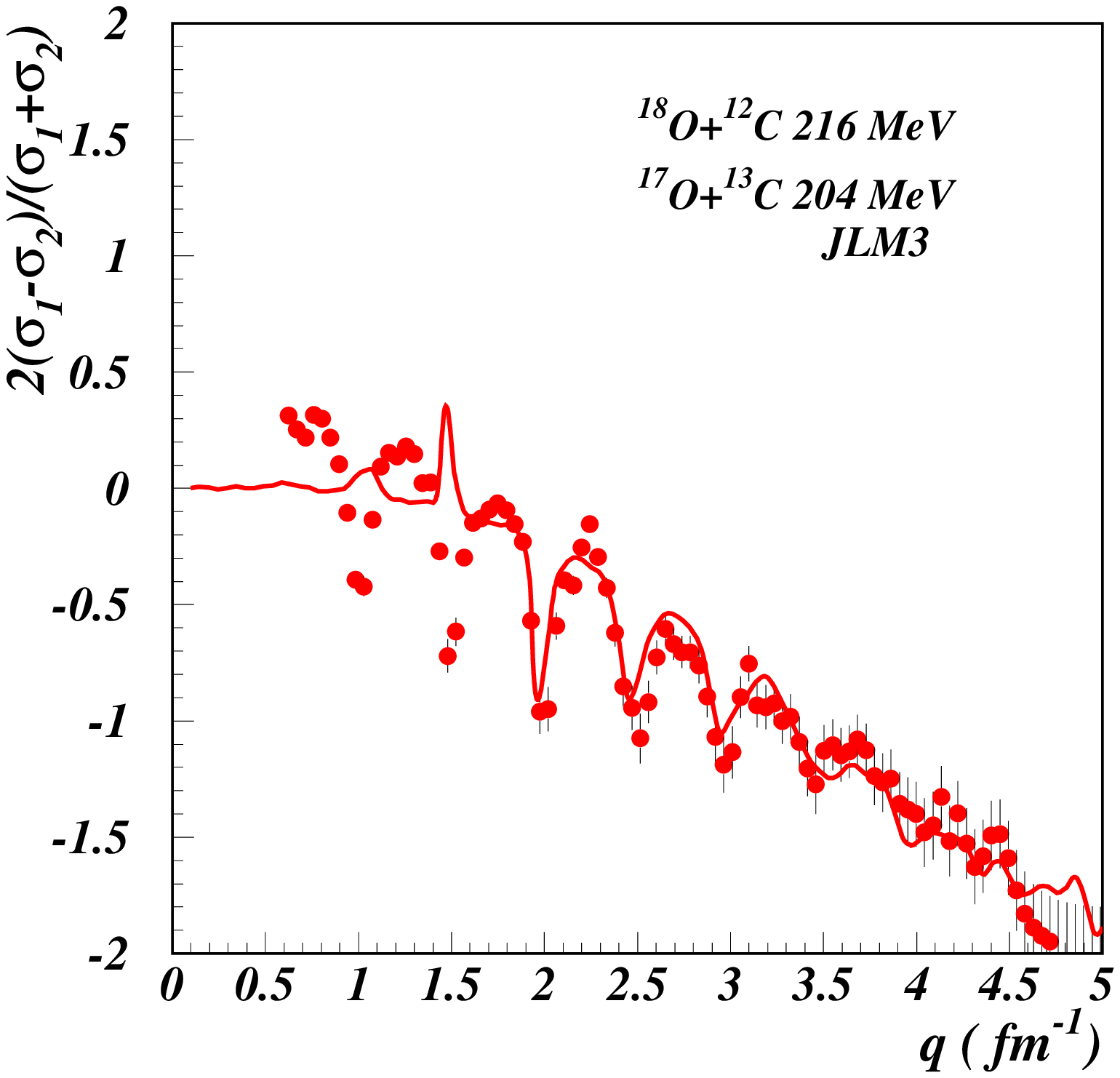}}{\caption{(Color online) Relative cross section for the
elastic scattering $^{18}$O+$^{12}$C and $^{17}$O+$^{13}$C reactions as a function of momentum transfer is compared
 with the JLM3 model.}\label{figtar9}}
\end{figure}

\begin{figure}[!ht]
{\includegraphics[trim= 1.0cm 1.0cm 1.1cm 2.0cm , clip=true, scale=0.5]{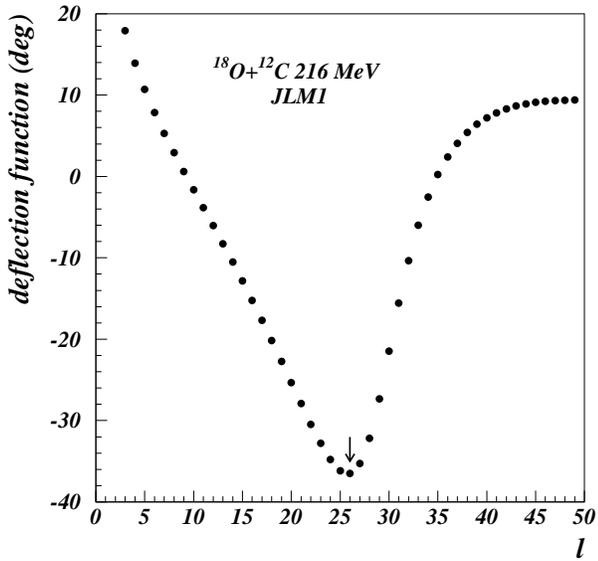}}{\caption{Classical
deflection function for the WS potential equivalent to JLM1. The rainbow angle is $\theta_R=36^\circ$. The
entire measured angular range is illuminated.}\label{figtar10}}
\end{figure}

\begin{figure}[!ht]
{\includegraphics[trim= 1.0cm 1.0cm 1.1cm 1.0cm , clip=true, scale=0.5]{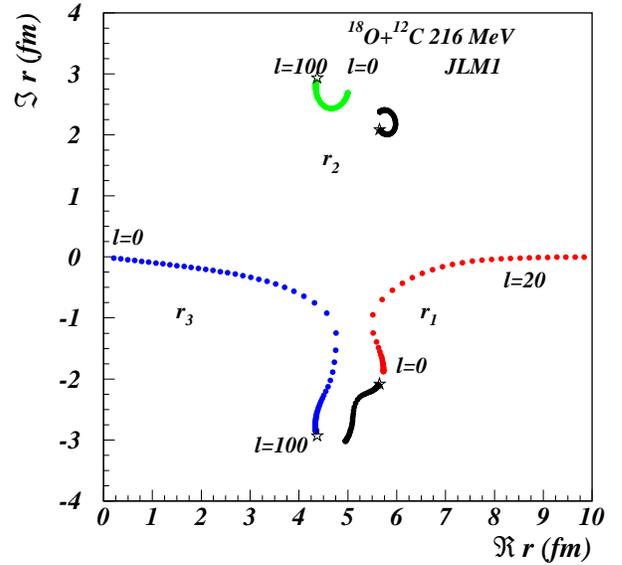}}{\caption{(Color online) Complex
turning points for the WS potential equivalent to JLM1. The stars denote the complex poles of the potential.}\label{figtar11}}
\end{figure}

\begin{figure}[!ht]
{\includegraphics[trim= 1.0cm 1.0cm 1.1cm 1.0cm , clip=true, scale=0.5]{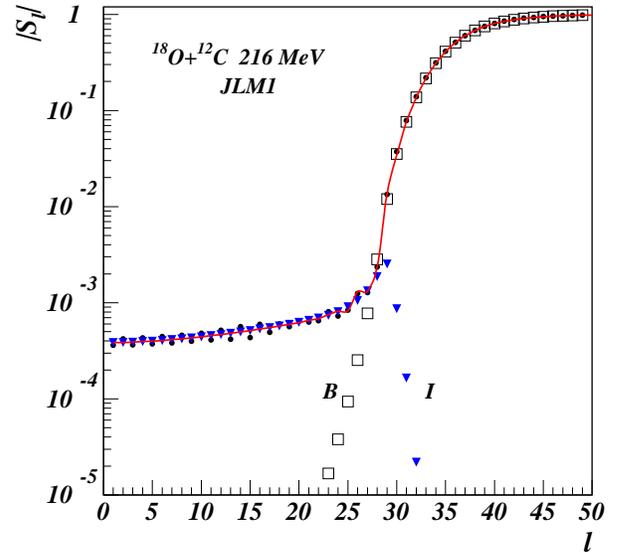}}{\caption{(Color online) Semiclassical
absorption profile for the WS potential equivalent to JLM1. The barrier and internal barrier components of the S-matrix are
shown by open squares and triangles respectively. The barrier component is typical for strong absorption. The black dots denote the exact quantum result for the same potential and the line
is a spline interpolation of the WKB S-matrix.}\label{figtar12}}
\end{figure}

\begin{figure}[!ht]
{\includegraphics[trim= 1.0cm 1.0cm 1.1cm 1.0cm , clip=true, scale=0.5]{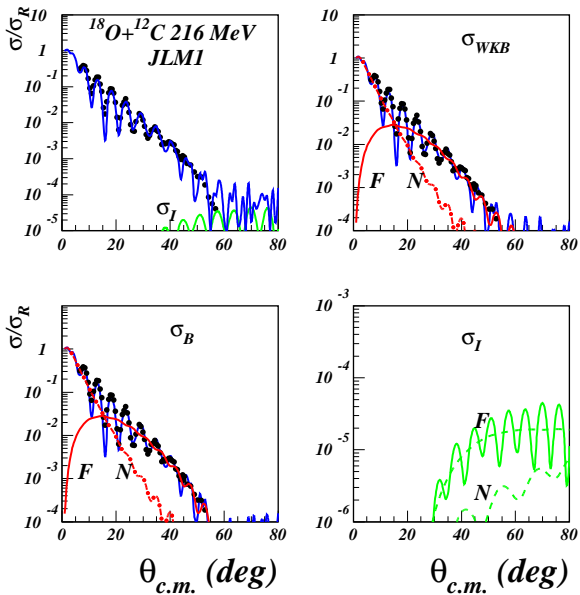}}{\caption{(Color online) Semiclassical
calculation of the cross section based on the WS potential equivalent to JLM1. The WKB scattering amplitude is further
decomposed into barrier ($\sigma_B$) and internal barrier ($\sigma_I$) components. The internal barrier component is
negligible small in the measured angular range. The reaction is peripheral.}\label{figtar13}}
\end{figure}

\begin{figure}[!ht]
{\includegraphics[trim= 1.0cm 1.0cm 1.1cm 1.0cm , clip=true, scale=0.5]{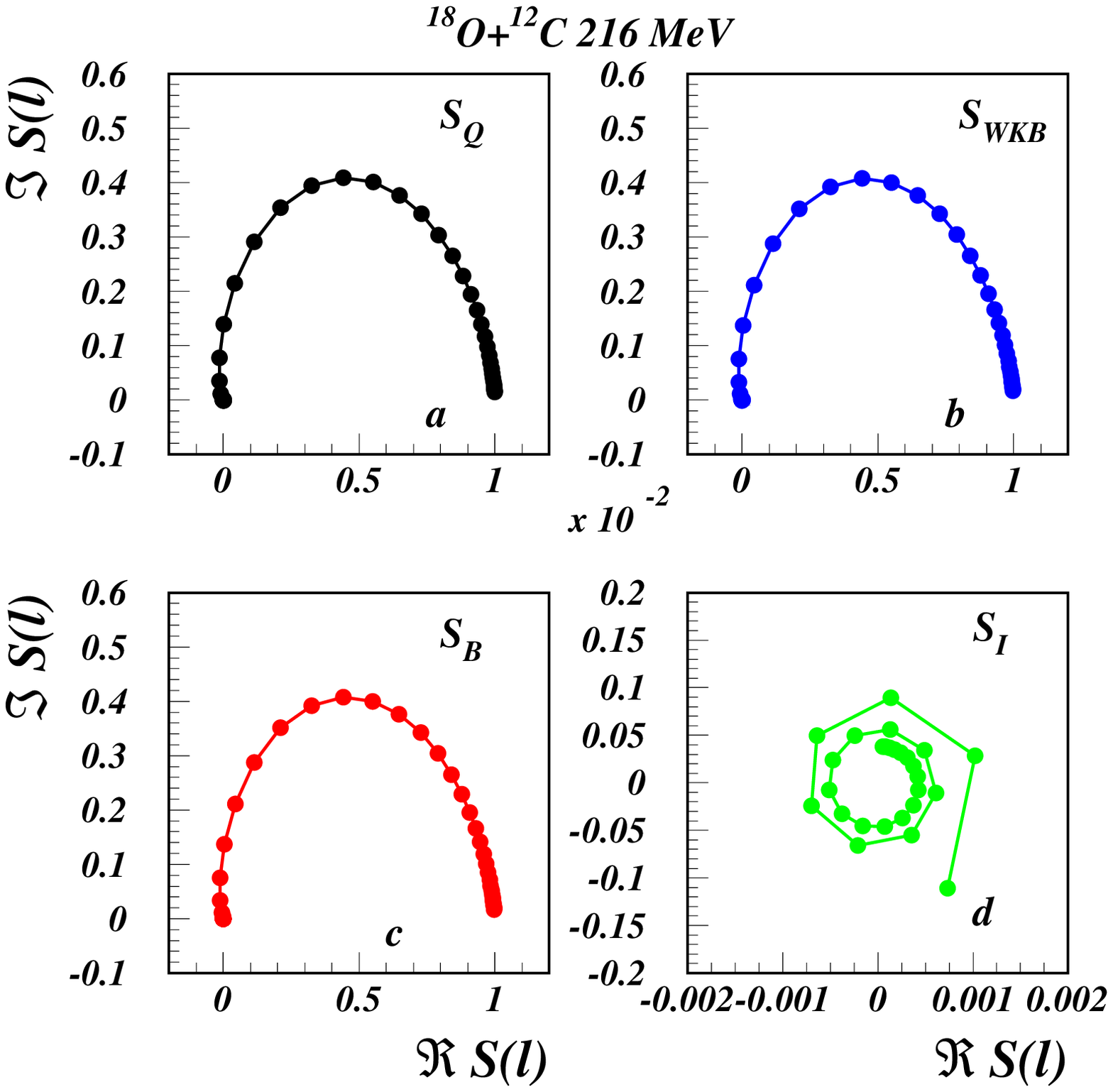}}{\caption{(Color online) Argand diagram for the
semiclassical S-matrix based on  the WS potential equivalent to JLM1. The barrier trajectory (panel c) is identical to
the exact quantum result (panel a). The small internal barrier component (panel d) shows a hint of an orbiting
effect or the presence of Regge poles, but these are too far from the real axis to have noticeable effect in
the total cross section.}\label{figtar14}}
\end{figure}

%\begin{figure}[!ht]
%{\includegraphics[width=0.9\textwidth]{inelast17.eps}}{\caption{(Color online) Inelastic cross section
%to $5/2^+ (3.84 MeV)$  state in $^{17}$O. The DWBA calculation is based on the T1 %potential.}\label{inelasto17}}
%\end{figure}

Above $\beta_i=(A_i-1)/A_i$ are recoil corrections, $\hat j_1(x)=3j_1(x)/x$ and $j_{0,1}$ are spherical
Bessel functions. The local Fermi momenta $k_f$ are evaluated in an
extended Thomas-Fermi approximation \cite{thomas}. We have explored also the
extended Slater approximation for the mixed densities of Campi and Bouyssy
\cite{campi} but did not obtained substantial improvements over the usual
Slater approximation. The local momentum for the relative motion is given by,

\be
K^2(R)=\frac{2\mu}{\hbar^2}(E_{c.m.}-U_D(R)-U_L(R))
\label{eqf10}
\ee
where $U_D$ is the total direct component of the potential including the Coulomb term. In Eq.
(\ref{eqf10}) we assumed  a purely real  local momentum of the relative motion  since the
absorptive component of the OMP is small compared with the real part. The effective mass correction \cite{negele},
$\frac{\mu^{\star}}{\mu}=1-\frac{\partial U}{\partial E}$ is of the order of a few percent for our
systems and is absorbed in the renormalization parameter $N_W$. Some tens of
iterations are needed to solve Eq. (\ref{eqf9}) in order to obtain a precision of
10$^{-7}$ in the entire radial range ( $R_{max}=25$ fm ). Calculations with finite range model are dubbed
M3YFR.
%taiat

 Neglecting the spin-orbit
component, the Gogny NN effective interaction can be expressed
as a sum of a central, finite range term and a zero range density dependent term,
\ba
v(\vec r_{12})=\sum_{i=1}^2(W_i+B_iP_{\sigma}-H_iP_{\tau}-M_iP_{\sigma}P_{\tau})e^{-\frac{r_{12}^2}{\mu_i^2}}\\ \nonumber
+t_3(1+P_{\sigma})\rho^{\alpha}(\vec R_{12})\delta(\vec r_{12})
\label{eqf11}
\ea
where $\vec r_{12}=\vec r_1-\vec r_2$ , $\vec R_{12}=(\vec r_1+\vec r_2)/2$ and
standard notations have been used for parameter strengths and  spin-isospin
exchange operators. The
strengths parameters and the ranges are taken from \cite{gogny}. The isoscalar and isovector components
of the effective interaction are constructed in the standard
way. The interest in this
interaction resides in its excellent description (at the HF level) of the saturation properties of
the nuclear matter in line with modern estimation from the isoscalar giant
monopole \cite{young} or dipole resonance \cite{colo} studies. Antisymmetrization of the
density dependent term is trivial, so that the sum of direct and exchange term
reads,
\be
v_D^{\rho}(r_{12})+v_{ex}^{\rho}(r_{12})=\frac{3t_3}{4}\rho^{\alpha}\delta(\vec
r_{12})
\label{eqf12}
\ee
%aici
The local equivalent of the finite range knockon exchange is calculated with Eq.
(\ref{eqf9}).  Two approximations were used for the overlap density,
\be
\rho=(\rho_1(r_1)\rho_2(r_2))^{1/2}
\label{eqf13}
\ee
and
\be
\rho=\frac{1}{2}(\rho_1(r_1)+\rho_2(r_2))
\label{eqf14}
\ee
The first approximation Eq.(\ref{eqf13}) has the merit that the overlap density goes to zero when one of the interacting
nucleons is far from the bulk. In Eq. \ref{eqf14} a factor 1/2 was introduced such
as the overlap density does not exceeds the equilibrium density for
normal nuclear matter. At large density overlaps, the fusion and other inelastic
processes are dominant and the elastic scattering amplitude is negligible small.
The calculated OM potentials are dubbed GOGNY1 and GOGNY3 respectively. Both definitions
represent crude approximations of the overlap density but are widely used
in the estimation of the density dependence
effects in the folding model.

We further examine the density dependence effects by using
 the nuclear
matter approach of Jeukenne, Lejeune and Mahaux (JLM) \cite{jeuken} which
incorporates a complex, energy and density dependent parametrization of the
NN effective interaction obtained in a Brueckner Hartree-Fock approximation
from the Reid soft core NN potential. The systematic study \cite{trache00} of
the elastic scattering between $p$-shell nuclei at energies
around 10 MeV/nucleon leads to the surprising result that on average, the
imaginary part of the folded JLM potential was perfectly adequate to
describe such reactions and did not need any renormalization ($N_{W}=1.00\pm
0.09$), while the real component needed a substantial renormalization, in line
with other effective interactions used in folding models. We examine here to which
extent this feature is conserved for tightly bound nuclei in the
$d$ shell in the presence of a small neutron excess.
Exchange effects are included in this model at the level of N-target
interaction. Calculations with this model are dubbed JLM1 and JLM3,
depending on
which definition we
use for the overlap density (eqs.(\ref{eqf13}) and (\ref{eqf14}) respectively).

A grid search on the real volume integral
reveals a unique solution for all six versions of thee effective interaction, see   Table \ref{tableo18c12216mevfolding} and Figs
\ref{figtar5}, \ref{figtar6}, \ref{figtar7} and \ref{figtar8}. The folding model
validates only the  solution with the lowest real
volume integral found with the WS parametrization. Averaging over all six
folding calculations, we find $J_V=167\pm 9$ MeV fm$^3$ for $^{18}$O
and $J_V=194\pm 5$ MeV fm$^3$ for $^{17}$O and so the interaction of $^{17}$O is
slightly more refractive. Again imaginary volume
integrals are quite small pointing to a some transparency of the potential.
Correction due to the finite range effects are quite large, of the order of
$\Delta R\approx 0.5$ fm for the real potential and much larger for the
imaginary potential. The folding calculation reproduces perfectly the
diffractive pattern at forward angles and the Fraunhofer F/N crossover produces
always an interference maximum. Beyond the cross-over the far-side component
decays quite smoothly and shows some glory effects at $\theta>60^{\circ}$.

More information one can extract from Fig. \ref{figtar9} where we plot the
spectral gradient ( or relative cross section) \cite{auger},
\be
E(q)=2[\sigma_1(q)-\sigma_2(q)]/[\sigma_1(q)+\sigma_2(q)]
\label{eqfal}
\ee
where $\sigma_1$ and $\sigma_2$ denote the differential cross sections for
$^{18}$O and $^{17}$O and $q$ is the momentum transfer. The calculation is done
with the JLM3 model, since the Glauber model is questionable at this low energy.
 The pattern in
Fig. \ref{figtar9} confirms the diffractive character of our reactions and an
intricate interference effect arising from the variation in the radius of
optical model potential and its surface thickness. The disagreement at low momentum transfer
arises mostly from the lack of
long-range correlation in the HF+BCS model for open shell nuclei.
%taiat
At this point we want to make a comment on the role of the dynamic polarization
potential for nuclei with neutron excess over the closed shell. A close
examination of the results in Table \ref{tableo18c12216mevfolding} shows that we
have obtained consistent results for all effective interactions  used in
the folding model. Our results confirm the conjecture that one can extract from
the elastic scattering at best only the low momenta of the interaction (volume
integrals and $rms$ radii). Corrections in the range parameters are large
especially for the imaginary component of the optical potential.  We found substantial
renormalization for the real part of the optical
potential, on average $N_V=0.36\pm 0.05$ in line with the previous study \cite{trache00}.
This can be easily understood: the bare folding formfactor has a volume integral around
$J_V\approx 450$ MeV fm$^3$, while the data requires
precise values around 160-190 MeV fm$^3$. Noteworthy, the renormalization of the
imaginary component in the JLM model is again quite close to unity. Although the
density dependence in the GOGNY and JLM effective interactions is very
different, one cannot disentangle between the two models for the overlap density
based on the present
data, since both of them give identical results.

% end folding
%%%%%%%%%%%%%%%%%%%%%%%%%%%%%

\section{Semiclassical barrier and internal barrier amplitudes}

\label{semicl} Once we have established the main features of the average OM
potential, we turn now to study the reaction mechanism  using
semiclassical methods.

The semiclassical uniform approximation for the scattering amplitude of
Brink and Takigawa \cite{brink2} is well adapted to describe situations in
which the scattering is controlled by at most three active, isolated,
complex turning points. An approximate multireflection series expansion of
the scattering function can be obtained, the terms of which have the same
simple physical meaning as in the exact Debye expansion for the scattering
of light on a spherical well. The major interest in this theory comes from
the fact that it can give precious information on the response of a nuclear
system to the nuclear interior.
% fig 10

\begin{figure}[!ht]
{\includegraphics[ trim= 1.0cm 1.0cm 1.1cm 1.0cm , clip=true,scale=0.9]{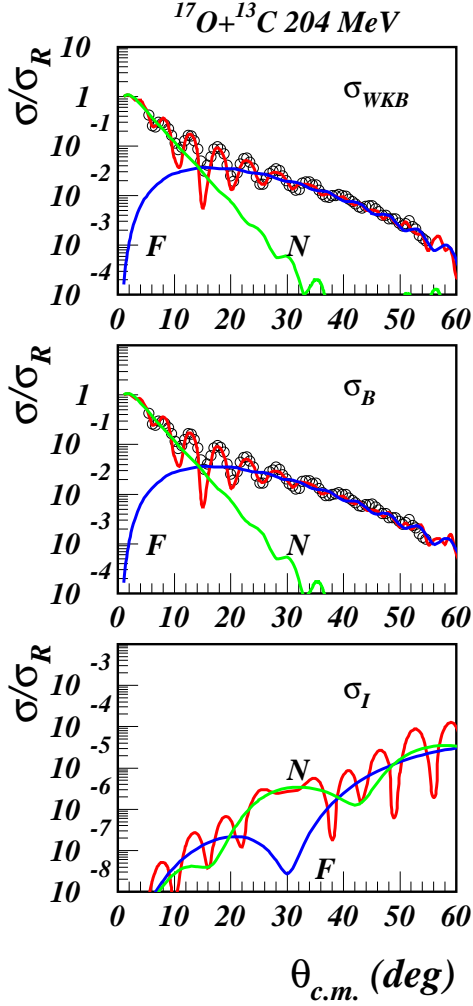}}{\caption{Semiclassical (WKB) calculation
of the cross section based on the T1 potential Table \ref{tableo18c12216mevws}. The barrier component match perfectly the data in the entire angular range, while the internal barrier
component is negligibly small.}\label{figtar15}}
\end{figure}
\begin{figure}[!ht]
{\includegraphics[width=0.9\textwidth]{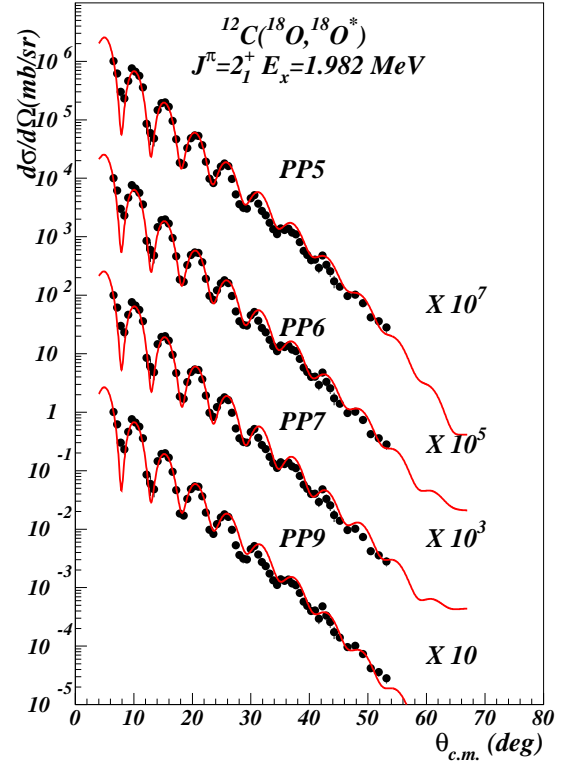}}{\caption{(Color online) Inelastic cross section
to $2_1^+$ (1.982 MeV)  state in $^{18}$O. The DWBA calculation is based on the
potentials in Table \ref{tableo18c12216mevws}.}\label{figtar16}}.
\end{figure}
We take as an example the potential PP9  in Table \ref{tableo18c12216mevws}
which is a WS phase
equivalent to the JLM1 optical potential. We discard the absorptive term and define the
effective potential as,
\begin{equation}
V_{eff}(r)=V(r)+\frac{\hbar ^{2}}{2\mu }\frac{\lambda ^{2}}{r^{2}},~~\lambda
=\ell +\frac{1}{2}  \label{eqsem1}
\end{equation}%
\noindent
where the Langer prescription has been used for the centrifugal term. This
guarantees the correct behavior of the semiclassical wave function at the
origin. Then we calculate the deflection function,
\begin{equation}
\Theta (\lambda )=\pi -2\int_{r_{1}}^{\infty }\frac{\sqrt{\frac{\hbar ^{2}}{%
2\mu }}\lambda dr}{r^{2}\sqrt{E_{c.m.}-V_{eff}}}  \label{eqsem2}
\end{equation}%
\noindent
where $r_{1}$ is the outer zero of the square root, i.e. the radius of
closest approach to the scatterer and $\mu $ is the reduced mass. Note that
with the replacement $\hbar \lambda =b\sqrt{2\mu E}$, Eq. \ref{eqsem2}
becomes identical with the classical deflection function $\Theta (b)$, where
$b$ is the impact parameter. The result is shown in Fig. \ref{figtar10}%
. The behavior of $\Theta (\lambda )$ is the one expected for an attractive
nuclear potential. The nuclear rainbow angle is $\theta_R\approx 36^{\circ}$. All the measured angular range is
classically illuminated and only a few points were measured in the dark side. This explains partially the ambiguities
found with the WS formfactors.

However this simple calculation does not  provide too much information about the interference effects of
the corresponding semiclassical trajectories.  Going into the complex $r$-plane we search
for complex turning points,
i.e. the complex roots of the quantity $E_{c.m.}-V_{eff}-iW$. This is an
intricate numerical problem, because, for a WS optical potential, the
turning points are located near the potential singularities and there are an
infinite number of such poles. The situation for integer angular momenta is
depicted in Fig. \ref{figtar11}. Active turning points are located near the poles of the real formfactor.
Inactive turning points are located quite far from the real axis and give
negligible small contribution to the total S-matrix. We observe
an ideal situation
with three, well isolated, turning points for each partial wave.
The multireflection expansion of the scattering function in the
Brink-Takigawa approach reads,
\begin{equation}
S_{WKB}(\ell )=\sum_{q=0}^{\infty }S_{q}(\ell )  \label{eqsem3}
\end{equation}%
where,
\begin{equation}
S_{0}(\ell )=\frac{\exp (2i\delta _{1}^{\ell })}{N(S_{21}/\pi )}
\label{eqsem4}
\end{equation}%
and for $q\not=0$,
\begin{equation}
S_{q}(\ell )=(-)^{q+1}\frac{\exp {[2i(qS_{32}+S_{21}+\delta _{1}^{\ell })]}}{%
N^{q+1}(S_{21}/\pi )}  \label{eqsem5}
\end{equation}%
In these equations $\delta _{1}^{\ell }$ is the WKB (complex) phase shift
corresponding to the turning point $r_{1}$, $N(z)$ is the barrier
penetrability factor,
\begin{equation}
N(z)=\frac{\sqrt{2\pi }}{\Gamma (z+\frac{1}{2})}\exp {(z\ln z-z)}
\label{eqsem6}
\end{equation}%
and $S_{ij}$ is the action integral calculated between turning points $r_{i}$
and $r_{j}$,
\begin{equation}
S_{ij}=\int_{r_{i}}^{r_{j}}dr\{\frac{2\mu }{\hbar ^{2}}[E_{c.m.}-V_{eff}-iW]%
\}^{1/2}  \label{eqsem7}
\end{equation}%

\begin{figure}[!ht]
{\includegraphics[width=0.9\textwidth]{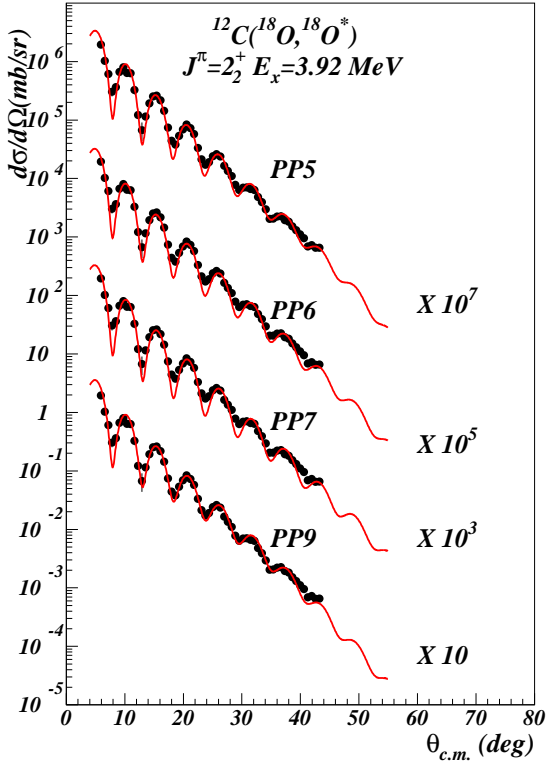}}{\caption{(Color online) Inelastic cross section
to $2_2^+$ (3.92 MeV)  state in $^{18}$O. The DWBA calculation is based on the  potentials in Table \ref{tableo18c12216mevws}.}\label{figtar17}}
\end{figure}

$S_{21}$ and $S_{32}$ are independent of the integration path provided they
lie on the first Riemann sheet and collision with potential poles is
avoided. Each term in Eq. \ref{eqsem3} has a simple physical interpretation.
The first term (the barrier term, denoted also $S_{B}$) retains
contributions from trajectories reflected at the barrier, not penetrating
the internal region. The $q$th term corresponds to trajectories refracted $q$
times in the nuclear interior with $q$-1 reflections at the barrier turning
point $r_{2}$. Summation of terms $q\geq 1$ can be recast into a single
term,
\begin{equation}
S_{I}=\frac{exp{[2i(S_{32}+S_{21}+\delta _{1}^{\ell })]}}{N(S_{21}/\pi )^{2}}%
\frac{1}{1+\exp {[2iS_{32}]/N(S_{21}/\pi )}}  \label{eqsem8}
\end{equation}%
and is known as the internal barrier scattering function. The last factor in Eq.
\ref{eqsem8}, the enhancement factor, is responsible for the multiple reflections
of the wave within the potential pocket.  When the
absorption in the nuclear interior is large, the enhancement  factor  reduces
to unity. Since
the semiclassical scattering function is decomposed
additively, $S_{WKB}=S_{B}+S_{I}$, the corresponding total scattering
amplitude is decomposed likewise as $f_{WKB}=f_{B}+f_{I}$ and conveniently
the corresponding barrier and internal barrier angular distributions are
calculated as $\sigma _{B,I}=|f_{B,I}|^{2}$, using the usual angular
momentum expansion of the amplitudes. % fig 12

\begin{figure}[!ht]
{\includegraphics[width=0.9\textwidth]{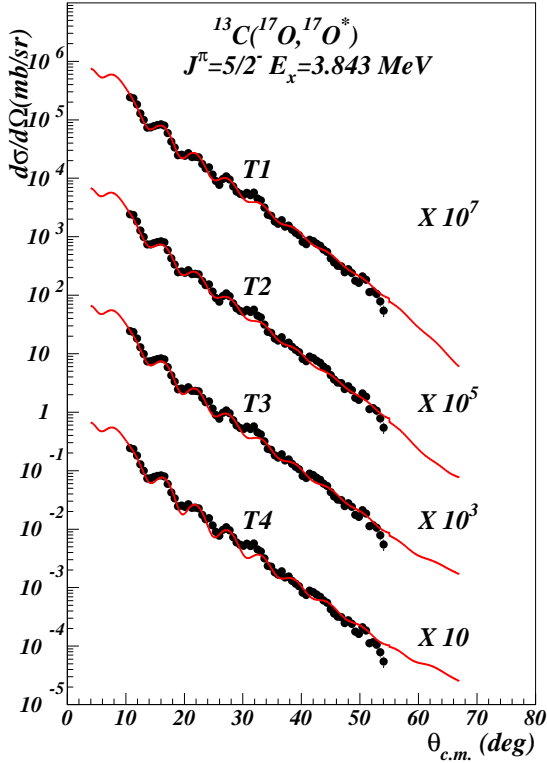}}{\caption{(Color online) Inelastic cross section
to $5/2^-$ (3.84 MeV)  state in $^{17}$O. The DWBA calculation is based on the  potentials in Table \ref{tableo18c12216mevws}.\label{figtar18}}}
\end{figure}
\begin{figure}[!ht]
{\includegraphics[width=0.9\textwidth]{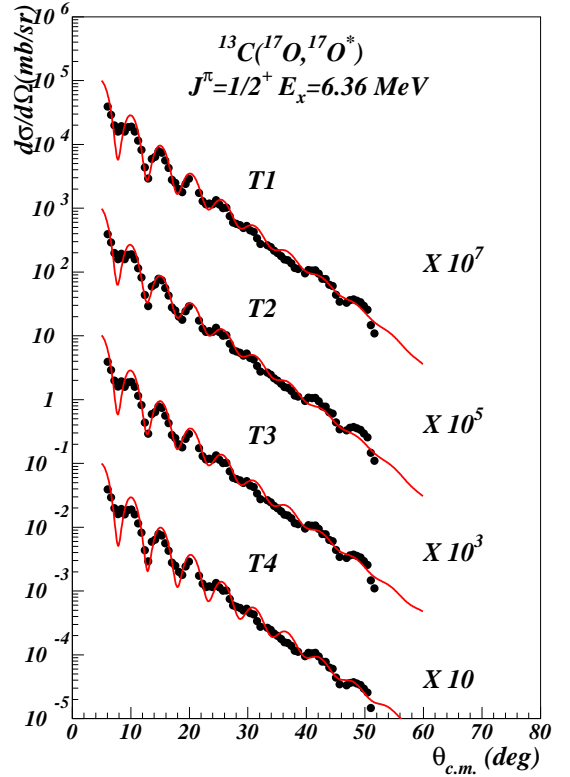}}{\caption{(Color online) Inelastic cross section
to $1/2^+$ (6.36 MeV)  state in $^{17}$O. The DWBA calculation is based on the  potentials in Table \ref{tableo18c12216mevws}.\label{figtar19}}}
\end{figure}

The poles of the semiclassical S-matrix are given by,
\begin{equation}
N(i\epsilon)+e^{2iS_{32}}=0~ ~ ; ~~ \epsilon=-\frac{i}{\pi}S_{21}
\label{eqsem9}
\end{equation}

Semiclassical Regge poles of Eq. \ref{eqsem9} are too far from the real axis
to have a noticeable influence on the total cross section. The accuracy of
the semiclassical calculation has been checked by comparing
the barrier and internal barrier absorption profiles with the exact
quantum-mechanical result in Fig. \ref{figtar12}. One observes
that the semiclassical B/I expansion is an $\mathit{exact}$ decomposition of
the quantum result. They are virtually identical at the scale of the figure.
The internal component gets significant values up to the grazing angular
momentum ($\ell _{g}$=36) and is negligible small beyond this value. The
barrier component resembles a strong absorption profile and this justifies
the interpretation that it corresponds to that part of the flux not
penetrating into the nuclear interior.
 Second, the B/I components are almost decoupled in the
angular momentum space and therefore they will contribute in different
angular ranges. % fig 13

Semiclassical cross sections are compared with the data in Fig. \ref{figtar13}. Better insight
into this technique is obtained by further decomposing the B/I components
into far and near (BF/BN and IF/IN) subcomponents. Clearly, the barrier
component dominates the entire measured angular range. Fraunhofer diffractive
oscillations appear as the result of BF and BN interference. At large
angles, the internal contribution is negligible and the reaction is peripheral.

The Argand diagrams corresponding to the B/I decomposition is displayed in Fig.
\ref{figtar14}.
The  barrier amplitude (panel c) is almost identical with the exact quantum result (panel a) while
the internal barrier component shows a nice orbiting effect, but the corresponding dynamical content
($S_I(\ell)$ is too small to have any sizeable effect in the total cross section.

A similar analysis was performed for the reaction $^{17}$O+$^{13}$C based on the WS potential, parameter set T1
Table \ref{tableo18c12216mevws}. Again we find that the WKB cross section is identical with the exact quantum result based on
the same potential. The barrier component match perfectly the data in the entire angular range, while the internal barrier
component gives negligible contribution, see Fig.\ref{figtar15}. Thus the peripherality character of our reactions is completely
demonstrated.

%%%%%%%%%%%%%%%%%%%%%%%%%
\section{Inelastic transitions}
We examine in this section the ability of our optical potentials to describe
the measured data for inelastic transitions
to selected states in $^{18}$O ( $J^{\pi}=2^+_1$, $E_x=1.982$  MeV, Fig. \ref{figtar16} and  $J^{\pi}=2^+_2$, $E_x=3.92$  MeV, Fig.
\ref{figtar17}) and two transitions in $^{17}$O ($J^{\pi}=\frac{5}{2}^-, E_x =3.843$ MeV, Fig. \ref{figtar18} and
$J^{\pi}=\frac{1}{2}^+, E_x =6.36$ MeV, Fig. \ref{figtar19}).

The pattern of our data shows a clear diffractive character since they obey
fairly well to the Blair phase rule \cite{blair} and therefore a standard
DWBA should be an appropriate approach.
The deformation table \cite{moller} indicates a quadrupole deformation $\beta_2=0.107$ for $^{18}$O. Other
systematics \cite{cdfe}
suggests puzzling results with quadrupole deformation ranging from 0.085 up to 0.339.
Since the  DWBA cross section
scales with $\beta_2^2$, we execute two calculations using $\beta_2=0.015$ and 0.35, chosen
rather arbitrary in the range
of suggested values. DWUCK4 and FRESCO give identical shapes for these two
values. We then scale the calculation to match the data. The scaled calculations are shown in Fig. \ref{figtar16}
and Fig. \ref{figtar17}.

The shape of the calculated cross section
is virtually identical for all the potentials at the scale of
the figure. This proves once again that our potentials are almost phase equivalent, small differences appearing
only at large angles much beyond the measured  angular range. Remarkably, the calculation with the PP9 parameter set, which
is a WS potential phase equivalent to JLM1 folding potential describes the data as well as the other parameter sets. The
situation is similar for the other folding potentials. Thus we have obtained a consistent description of both elastic and
inelastic cross section using a large palette of optical potentials.

The pattern of the measured transitions in $^{17}$O is quite different.  The cross section has
no forward maximum and
decays almost exponentially at large angles
with small amplitude wiggles. The experimental study by Cunsolo \etal  \cite
{cunsolo} using three particle transfer reaction showed that the low-lying negative
parity state in $^{17}$O, $J^{\pi}=\frac{5}{2}^-, Ex =3.843$ MeV is a member of  $^{16}$O $K^+$ $\alpha$-rotational
band coupled to $p_{1/2}$ neutron, and thus has
 a pure $4p-3h$ configuration. The state $J^{\pi}=\frac{1}{2}^+, E_x =6.36$ MeV,
located only 3 keV bellow the $\alpha$ threshold in
$^{17}$O is weakly populated in the reaction $^{13}$C($^6$Li,d)$^{17}$O \cite{kubono}. This state is astrophysically important since it is
considered  the main source of the $^{13}$C($\alpha$,n)$^{16}$O reaction rate uncertainty. According to Cunsolo \etal  \cite{angelo}
this state has
a dominant
$3p-2h$ structure and belongs to a $(sd)^3$,T=1/2 $^{17}$O rotational band. Repeating the procedure used for $^{18}$O we
obtain a satisfactory description of our data, see Figs. \ref{figtar18} and \ref{figtar19}.

%\begin{figure}[!ht]
%{\includegraphics[width=0.9\textwidth]{inelast18.eps}}{\caption{(Color online) Inelastic cross section
%to $2^+_1 (1.98 MeV)$ and $2^+_2 (3.92 MeV)$ states in $^{18}$O. The DWBA calculation is
%based on the PP5 potential.}\label{inelasto18}}
%\end{figure}

%\FloatBarrier
\section{Conclusions}
We have measured elastic scattering cross sections for $^{18}$O+$^{12}$C and
$^{17}$O+$^{13}$C at 12 MeV/nucleon as well as inelastic transition to
selected states in $^{18}$O$^*$ and $^{17}$O$^*$ in order to determine the optical
potentials needed to study the one neutron pickup reaction
$^{13}$C($^{17}$O,$^{18}$O)$^{12}$C. Optical potentials in both incoming and
outgoing channels were extracted from a standard analysis using Woods-Saxon
formfactors. Analysis in terms of semimicroscopic double folding formfactors,
using six different approximations for the NN effective interactions helped us to
eliminate the ambiguities found with WS potentials. Thus a unique solution emerged
from the analysis, which is quite surprising when the reaction mechanism is
dominated by strong absorption. We find that the neutron excess over the closed
shell leads to a less refractive interaction as compared with the closed
shell nucleus $^{16}$O. A detailed semiclassical analysis in terms of
barrier and internal barrier amplitudes of Brink and Takigawa demonstrated that
the flux penetrating the barrier has negligible contribution to the total cross
section, and thus the reactions are  peripheral. This provides a complete
justification for the use of ANC method to extract spectroscopic information from
the transfer reaction.
\vspace{1cm}
\begin{acknowledgments}
This work was supported in part by the U.S. Department of Energy
under Grant No. DE-FG02-93ER40773 and DE-FG52-06NA26207,
the NSF under Grant PHY-0852653, the Robert A. Welch Foundation
under Grant No. A-1082, and by CNCSIS (Romania) Grant PN-II-PCE-55/2011 and PN-II-ID-PCE-0299/2012.
We thanks V. Balanica for technical support and to drs. Roland Lombard
and Vlad Avrigeanu for correspondence.
\end{acknowledgments}

%\newpage %Just because of unusual number of tables stacked at end

\end{document}